# ENTANGLEMENT MANAGEMENT IN SPACE-BASED QUANTUM INFORMATION NETWORKS

Luca Paccard[1], Agathe Blaise[2], Fabrice Arnal[1], and Laurent de Forges de Parny[1]

[1]*Thales Alenia Space, 26, Avenue J-F Champollion, 31037, Toulouse, France*
[2]*Thales SIX GTS, 4 Avenue des Louvresses, 92230 Gennevilliers, France*

(contact: laurent.de-forges-de-parny@thalesaleniaspace.com)

**ABSTRACT**

With the evolution of quantum computing, quantum sensing and secure quantum communication protocols, the demand for global development of Quantum Information Networks (QIN) has become crucial. Satellites play an indispensable role in enabling connectivity across vast distances, transcending terrestrial limitations. In this article, we explore various ways in which satellites may be involved in the deployment of these novel networks from their integration into the network architecture to the challenges they face.

## 1. INTRODUCTION

The late 20th century was marked with the arrival of new technologies that manipulate and control quantum devices, now commonly referred to as the second quantum revolution. Manipulating quantum devices involves modifying their quantum states. Key properties such as state superposition, entanglement, the no-cloning theorem and the quantum teleportation, play significant roles in these technologies. Quantum Information Networks (QINs) aim to share quantum information, i.e. the amplitude of probability of the quantum states, over long distances with processes called entanglement swapping and quantum teleportation, which consumes quantum entanglement. This field attracts increasing interest, as QINs would allow unprecedented computing, sensing, and security capacities.

The fundamental resource of QIN is quantum entanglement. Quantum entanglement is a phenomenon in which a group of particles is generated, interacts, or shares spatial proximity in such a way that the quantum state of each particle of the group cannot be described independently of the state of the others, including when the particles are separated by a large distance, as demonstrated with the use of the Micius satellite [1][2]. A QIN is a network that allows generation, distribution and routing entanglement between many users, used for quantum use case purposes.

The distance between QIN users has a significant impact on communication performances, especially when considering global-scale distances. Quantum signals are inherently weak and cannot be amplified due to the no-cloning theorem, imposing limitations. Fiber links have practical distance constraints, while free space links enable longer communication distances. In order to extend connectivity to remote areas and allow for global connectivity, satellite nodes are mandatory, hence the involvement of Thales Alenia Space in the development of these novel systems.

The article highlights our motivation to develop space-based QIN, particularly the space segment required for a global connectivity. We discuss the different types of high-level architecture in which the satellite could be used as a node of the quantum network.

## 2. SPACE AND GROUND PATHS IN QUANTUM INFORMATION NETWORKS

We start the discussion with a very simple example where two users need to share quantum entanglement for quantum communication purpose. We assume that the two users are sufficiently far away (e.g., > 400 km) such that the quantum information network has two inter-connected segments: the ground and Space segments (Figure 2-1).

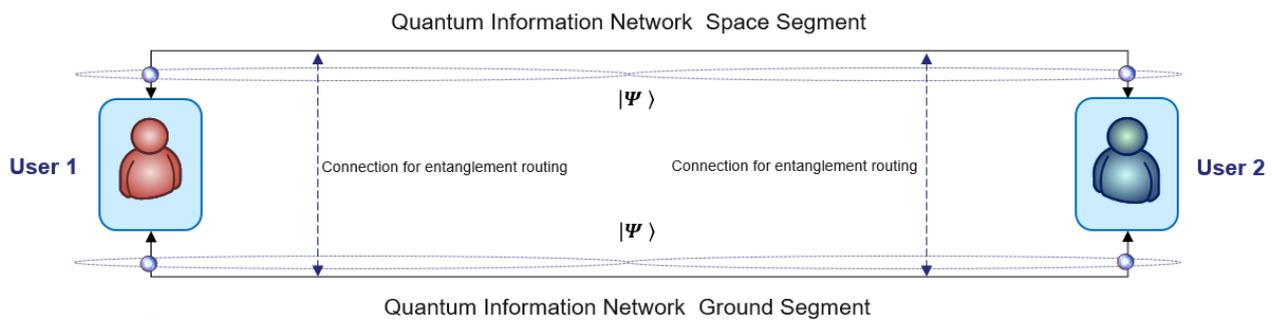

Figure 2-1: High-level quantum information network composed by a ground and a space segment, connecting two end-users. The two segments are assumed to be connected for routing purposes.

The two segments have the same function: to distribute entanglement to end-users. At the ground, connecting two users with an optical fiber, without quantum repeaters, implies inevitable photon losses that scale up exponentially with the transmission length. Ref [3] highlighted that at 1000 km, even with a perfect single-photon source of 10 GHz, ideal single photon detectors, and 0.2 dB/km fiber losses, one would detect only 0.3 photon on average per century! Although it is possible to amplify the signals 0 and 1 in classical communications, the situation is very different for quantum communications. Indeed, an unknown quantum superposition state cannot be noiselessly amplified. This is known as the quantum no-cloning theorem.

The extension of the quantum communication distance can be obtained, in principle, by plugging in a chain (space or ground) quantum repeater consisting of Bell state measurement (BSM) devices and that may include quantum memories, and (space or ground) entangled photon source modules (Figure 2-2 a)). Such a chain allows for changing the exponential scaling of the photon loss to a polynomial law as a function of the channel length. To this end, the entire channel between users 1 and 2 is divided into *N* segments such that each segment has an optimized direct transmission, i.e. an optimal signal-to-noise ratio (Figure 2-2 b)). The BSM and entangled photon sources are time-synchronized such that successive measurements of pairs of photons, each emitted from an entangled pair source, are performed at the BSM module. The two photons should arrive simultaneously through a beam splitter and must be quantum mechanically indistinguishable (same phase, same polarization, etc.). These measurements at the BSM module allow for entanglement swapping, a variant of quantum teleportation in which the particle to be teleported is itself part of an entangled pair [4][5]. Entanglement swapping provides an ideal method for entangling remote particles without a direct interaction.

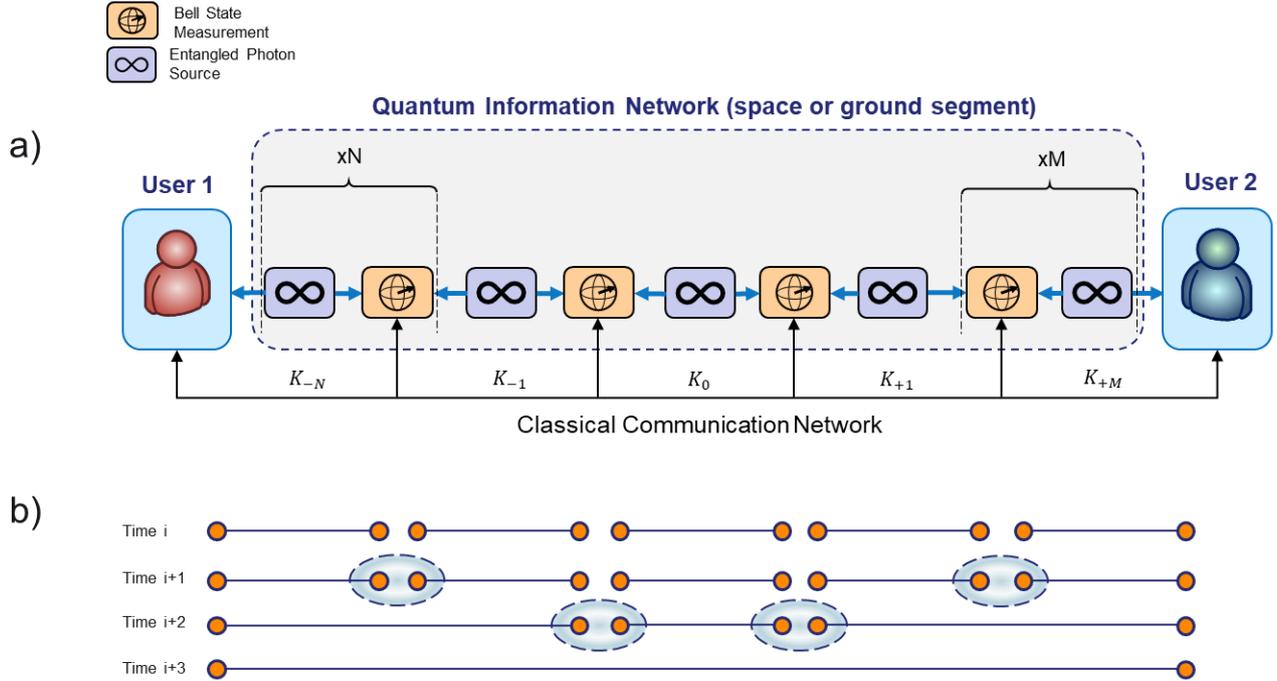

Figure 2-2: a) High-level QIN architecture, connecting two remote users. The quantum chain is composed by time-synchronized Bell state measurement modules and entangled photon source modules. The QIN is embedded in a classical communication network (e.g. Internet). b) Establishment of end-to-end entanglement, in four time steps, involving four Bell state measurements: two at time $i+1$ and two at time $i+2$. The end users share entanglement at time $i+3$.

The Bell state measurement modules and entangled photon source modules can be located either at the ground or on-board a satellite, depending on the distance between the two remote users. The role of the satellite and its integration into the QIN can have different variants, as discussed below.

### 3. SATELLITE INTEGRATION INTO QUANTUM INFORMATION NETWORK

As discussed in Section 2, a QIN is a network involving different quantum devices such as entangled photon sources, quantum memories and Bell state measurement for entanglement swapping. We also highlighted the requirement of the use of the satellite for extending the network scale. The question is now to define what is integrated on-board the satellite. To this purpose, we classify two types of architectures, those with and without Quantum Optical Inter Satellite Link (QOISL).

**Architectures without quantum optical inter satellite link:**

This first class of architecture assumes that satellites cannot exchange qubits between each other. Therefore, the satellite only exchanges qubits with optical ground stations, through downlink or uplink.

Scenario 1: Entangled photon source on-board the satellite

The first natural scenario considers that the satellite is equipped with an entangled photon source in downlink configuration (Figure 3-1). The entangled photon pairs are sent to two users. Users are equipped with quantum memories for the storage of qubits. The scenario depicted in Figure 3-1, b) – removing quantum memories – is very similar to the Micius one for entanglement distribution from low Earth orbit [1], and its extension for entanglement-based quantum key distribution [2]. An alternative scenario considers the satellite as a user itself, see Figure 3-1 a). One can assume a quantum memory on-board the satellite for the storage of qubits over time. At the moment, connecting a quantum memory on a ground station for storing space qubits belongs to the prospective research due to the very low technical readiness level of quantum memories. Embarking a space quantum memory is unrealistic at the moment.

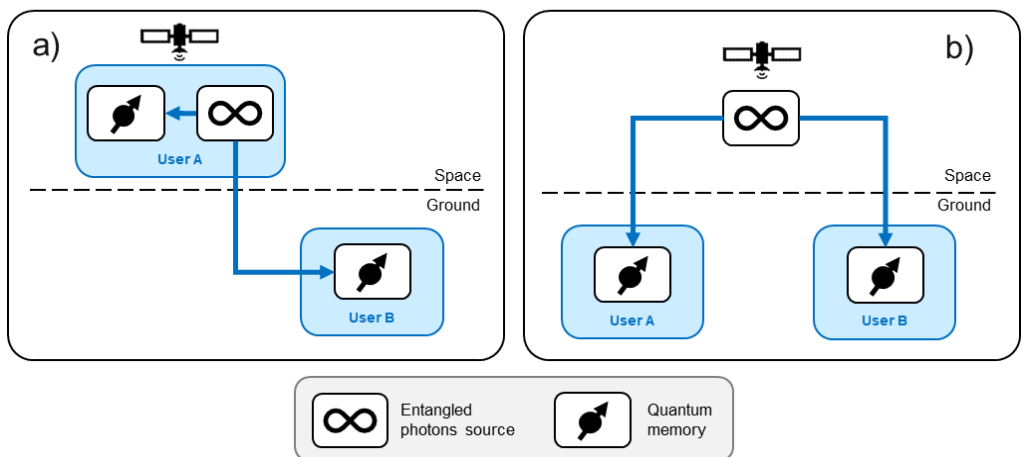

Figure 3-1: Scenario 1 a) and b) – The satellite is equipped with an entangled photon source in downlink configuration. a) The entangled photon pairs are sent to one user at the ground and to a device (e.g., a quantum memory) inside the satellite. b) The entangled photon pairs are sent to two users at the ground. The users can be equipped with quantum memories for the storage of the qubits.

Scenario 1 b) has a limited range of distance between the two users, constraint by the dual visibility of the two stations at the same time by the satellite. Indeed, as the distance between ground users increases, the satellite's altitude must be higher to ensure dual visibility. However, the higher the satellite's altitude, the weaker the signal received on the ground, due to a reduction in the link budget. Typically, the maximum altitude should be around 1000 km for conserving a credible level of service with realistic size of ground and space optical terminals. In conclusion, scenario 1 b) will be possible only for a limited range between the two users and other strategies should be investigated for greater distances between users.

Scenario 1 c) shows in Figure 3-2 a candidate architecture for distant users connected by using one quantum repeater node at ground and one satellite with an entangled photon source on-board. The quantum repeater extends the communication range between two users, allowing user A to connect without requiring direct, real-time visibility of the satellite. It also enables users without telescope-equipped ground stations to receive photons from the satellite, facilitating long-range entanglement. Without quantum memory, this architecture requires a near-perfect time-synchronization (~ 100 ps) at the repeater node in the measurement of the two photons to perform a BSM – one photon coming from the satellite and the second one coming from the entangled photons source located at the ground. The two photons should be indistinguishable for having a successful entanglement. This architecture also illustrates the use quantum memories at the user end-nodes for storage and at the Bell state measurement device for time-synchronization purposes, as well as to allow photons to be received at different times before performing a BSM. A possible extension of this architecture is shown in Figure 3-2 d) with one quantum repeater at ground and two satellites. This architecture further extends the range between users, but introduces greater complexity, as the satellites must pass over the quantum repeater at roughly the same time, depending on the constraints of quantum memory storage duration.

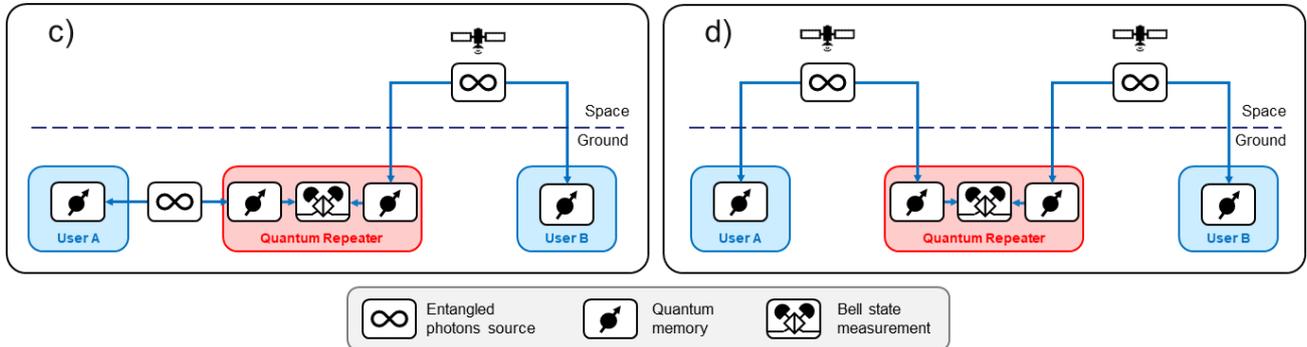

Figure 3-2: Scenario 1 c) and d) for which entangled photons cannot be distributed directly to the users – The satellite is still equipped with an entangled photons source in downlink configuration, as previously. c) The entangled photon pairs are sent to user B at the ground and to a Bell state measurement node, acting as a quantum repeater connected to user A. The users and the Bell state measurement are equipped with quantum memories for the storage of qubits. d) A variant from c) with two satellites for expanding again the range of the link.

Scenario 2: Bell state measurement on-board the satellite

Another architectural approach involves uplink configurations, where the satellite, equipped with a Bell state measurement device (quantum repeater) and quantum memories, receives qubits on-board. Compared to downlink, uplink configurations are less efficient due to the point-ahead and anisoplanatic angles caused by atmospheric turbulence, which are difficult to correct even with highly efficient adaptive optics. This setup requires on-board quantum memories and Bell state measurement devices, resulting in significantly more complex size, weight, and power (SWaP) demands than systems with on-board sources, due to the relative immaturity of components like quantum memories. This architecture is referred to as Measurement Device Independent (MDI) [6].

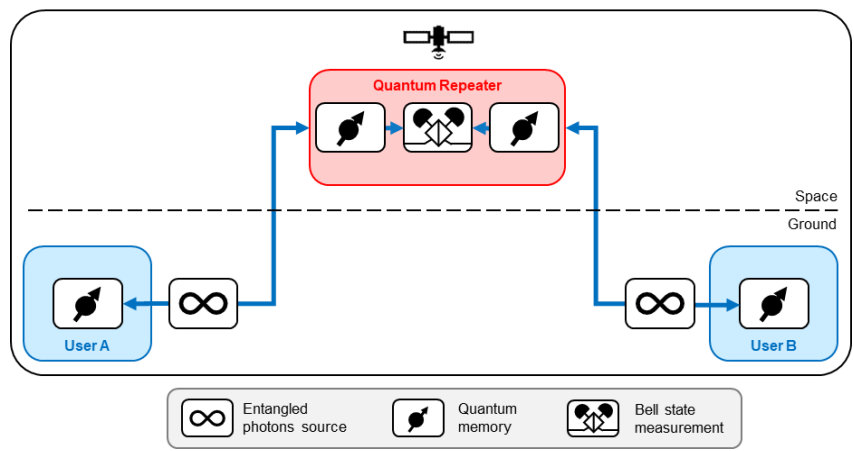

Figure 3-3: Scenario 2 – The satellite, equipped with a Bell state measurement (quantum repeater) with two quantum memories, receives two independent photons, each coming from an entangled photon source located at the ground. The entanglement swapping is performed when the two photons are measured at the Bell state measurement on-board. After this step, users A and B receive entangled photons and share entanglement.

**Architectures with quantum optical inter satellite links:**

In this section we consider quantum links between satellites and discuss all the new possibilities. What is particularly interesting is that optical inter-satellite links already exist for classical communications and can be adapted for quantum communications with relatively small size of optical terminals, compared to those used for space-to-ground links. Also, QOISL are not affected by atmospheric impairment, which is a great advantage for the losses in the channel.

A key design choice for QOISL-based QINs lies in the satellite architecture. The space segment can consist of different types of satellites, where some function as sources of entangled photons, while others serve as quantum repeaters. Alternatively, a more flexible approach would involve designing hybrid satellites capable of switching between roles as sources or repeaters, depending on the operational needs at any given moment. The same on-board terminals could be used for emission and reception, and we assume here a separated payload, i.e., no interaction between the on-board entangled photon source and the on-board Bell state measurement.

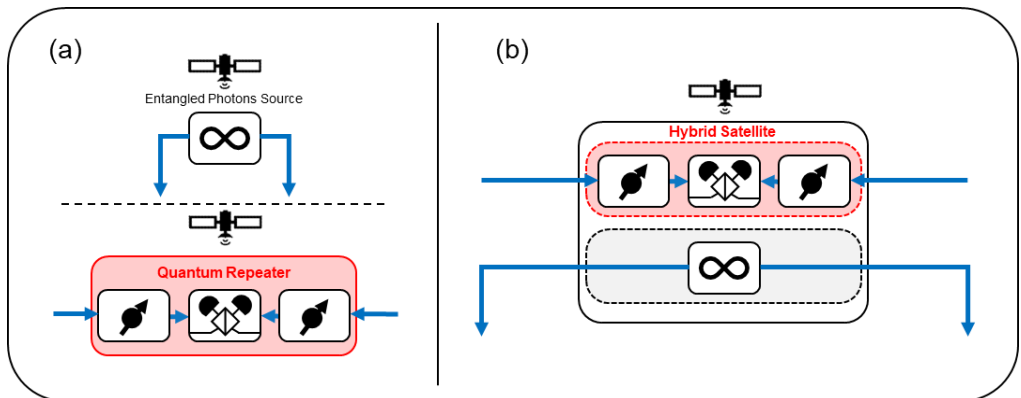

Figure 3-4 : Satellite architecture with (a) two types of satellites or (b) one type of hybrid satellite

Each architecture presents Pros and Cons detailed in Table 1.

**Table 1 : Trade-off of the satellites architecture**

| Architecture | PROS | CONS |
|---|---|---|
| **Specific satellite** | • Relatively low size, weight and power<br>• Low design complexity (compared to hybrid satellites) | • Complex constellation layout: two adjacent repeater satellites are useless without a source satellite in-between<br>• Must be compatible with non-OQISL uses: if downlink configuration, no entanglement distribution possible if the only satellite available to connect two ground stations is a quantum repeater |

| Architecture | PROS | CONS |
|---|---|---|
| **Hybrid satellite** | • Dynamic adaptation to network needs<br>• No specific arrangement of satellites in the constellation<br>• May require less satellites in the constellation | • Higher complexity to integrate both functions onboard a single satellite<br>• Higher size, weight, power and cost |

Scenario 3: Source satellite linked with BSM satellite

Combining specific satellites and mixing it with QOISL, we obtain scenario 3 where entangled photon sources on-board satellites are connected with space quantum links to a Bell state measurement on-board another satellite (Figure 3-5).

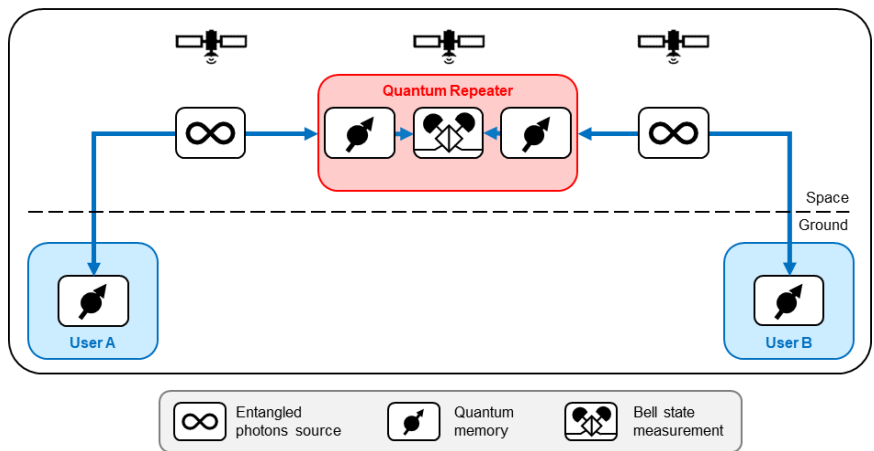

Figure 3-5: Scenario 3 a) – Two satellites, each with an entangled photons source on-board, are connected through two space quantum links (QOISL) to another relay satellite with a BSM on-board.

Alternatively, scenario 3 a) can be simplified with only one QOISL by considering only one quantum source satellite, the other one being located at the ground, leading to scenario 3 b) in Figure 3-6. However, this scenario re-introduces uplink configurations. Both scenarios 3 a) and 3 b) can be imagined with the architecture consisting of two different satellites or with hybrid satellites.

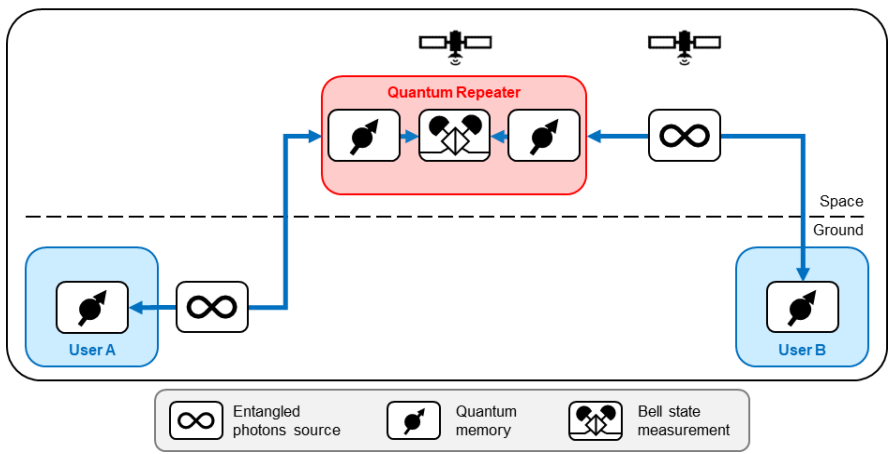

Figure 3-6: Scenario 3 b) – One quantum source satellite and one ground entangled photon source are interacting with one space relay node. Simplified version of scenario 3 a), here with only one QOISL.

Scenario 4: Source satellite linked with a hybrid satellite

The last interesting scenario is the case of a hybrid satellite interacting with a user and an entangled photon source on-board a satellite, thus involving one QOISL (Figure 3-7). This scenario considers a hybrid satellite with a direct interaction between the on-board entangled photon source and the on-board quantum repeater. Scenario 4 allows for a reception of qubits on one side of the satellite and an emission of the qubits on the other side.

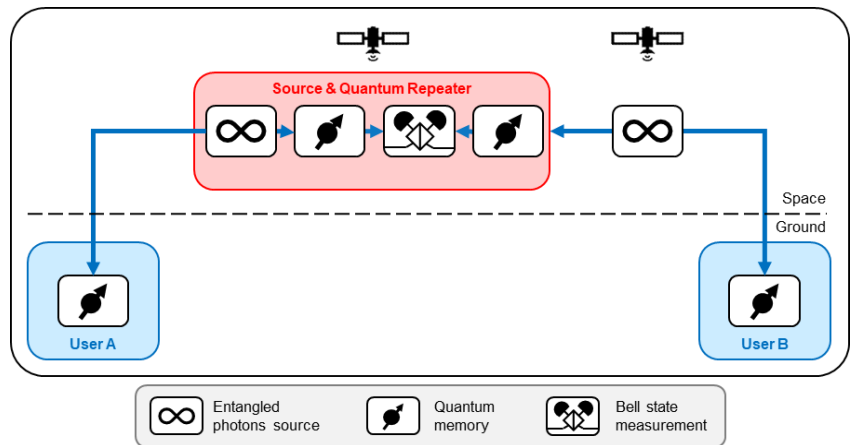

Figure 3-7: Scenario 4 – A hybrid satellite, composed of an entangled photon source plugged to a quantum repeater module (quantum memories and BSM), is connected through one QOISL to an entangled photon source on-board another satellite. Here, only one QOISL is involved.

**Comparison of the architectures:**

We have presented four architecture scenarios involving satellites in QINs. It was practical to class them in two categories, depending on the use or not of quantum optical inter-satellite links. We now propose to trade-off these architectures in order to highlight the Pros/Cons for each solution. For our trade-off, we consider the technological complexity for the realization of the payload, the size and number of telescopes, the number of links and their directivity, the pointing requirements, the use of space links or not, the use of (space) quantum memory and single photon detectors or not. The interest, in term of functionalities enabling, is also a key driver.

**Table 2: Trade-off of the QIN satellites scenarios**

| Scenario | Description | Pros | Cons | Complexity/5 | Interest/5 |
|---|---|---|---|---|---|
| 1 b) | Entangled photon source on-board the satellite | <ul><li>Downlink has lower losses compared to uplink</li><li>Photonic equipment feasible (see [1][2])</li><li>Medium technological complexity</li><li>No space quantum memory (for scenario 1 b))</li><li>No BSM</li><li>Space single photon detectors are not mandatory (except maybe for source health monitoring)</li></ul> | <ul><li>Requires two large telescopes on-board the satellite</li><li>Requires pointing toward two ground stations simultaneously</li><li>Involve two space-to-ground links</li><li>Involve space quantum memory for scenario 1 a)</li></ul> | 2 | 5 |
| 2 | Bell state measurement on-board the satellite | <ul><li>Allows the entanglement of remote photons at large distances</li></ul> | <ul><li>Requires two large receiving telescopes on-board the satellite</li><li>Requires pointing toward two ground stations simultaneously</li><li>Involve two ground-to-space links</li><li>Uplink has higher losses compared to downlink</li><li>Photonic equipment never realized in space</li><li>Higher technological complexity: involves space quantum memory and stringent time-synchronization</li><li>Involves space single photon detectors</li></ul> | 4 | 3 |
| 3 a) | Source satellites linked with BSM satellite | <ul><li>QOISL allows a better link budget than space-to-ground link (with atmospheric impairments)</li><li>QOISL allows for an extension into a satellite constellation</li><li>QOISL telescopes could be smaller than space-to-ground telescope</li><li>Reuse of existing OISL telescopes</li></ul> | <ul><li>Space sources & space BSM to be synchronized</li><li>Photonic equipment never realized, for the BSM payload</li><li>Higher technological complexity for the BSM module</li><li>Involves space quantum memory and stringent time-synchronization, for BSM</li><li>Involves space single photon detectors</li><li>Requires pointing toward a ground stations and a satellite simultaneously, or two satellites</li></ul> | 4 | 5 |

| Scenario | Description | Pros | Cons | Complexity/5 | Interest/5 |
|---|---|---|---|---|---|
| 3 b) | Source satellite linked with BSM satellite, other source at the ground | <ul><li>QOISL allows a better link budget than space-to-ground link (with atmospheric impairments)</li><li>QOISL allows for an extension into a satellite constellation</li><li>QOISL telescopes could be smaller than space-to-ground telescope</li><li>Reuse of existing OISL telescopes</li><li>Only one QOISL to be managed in this configuration</li></ul> | <ul><li>Space and ground sources & space BSM to be synchronized</li><li>Photonic equipment never realized, for the BSM payload</li><li>Higher technological complexity for the BSM module</li><li>Involves space quantum memory and stringent time-synchronization, for BSM</li><li>Involves space single photon detectors</li><li>Requires pointing toward a ground stations and a satellite simultaneously</li><li>Involves an uplink. Uplink has higher losses compared to downlink</li></ul> | 4 | 4 |
| 4 | Source satellite linked with a hybrid satellite | <ul><li>QOISL allows a better link budget than space-to-ground link (with atmospheric impairments)</li><li>QOISL allows for an extension into a satellite constellation</li><li>QOISL telescopes could be smaller than space-to-ground telescope</li><li>Reuse of existing OISL telescopes</li><li>Only one QOISL to be managed in this configuration</li><li>No uplink involved</li></ul> | <ul><li>Space source & BSM to be synchronized</li><li>Photonic equipment never realized, for the BSM payload</li><li>Higher technological complexity for the BSM module</li><li>Involves space quantum memory and stringent time-synchronization, for BSM</li><li>Involves space single photon detectors</li><li>Requires pointing toward a ground stations and a satellite simultaneously</li></ul> | 5 | 2 |

In conclusion, only Scenario 1 b) "Entangled photon source on-board the satellite" seems realistic in the near future. The use of uplinks and on-board Bell state measurements with space quantum memories are of higher technical level and require a significant R&D effort. However, certain long-term scenarios involving QOISL, such as Scenario 3 a) which has high-interest rank, could offer significant advantages for extending the range of quantum networks.

## 4. CONCLUSIONS

This article discuss a quantum communication topic of growing interest, the Quantum Information Network (QIN) which is much more general than QKD network applications, and for which the satellite has a crucial role to play. Quantum information networks are designed to connect quantum computers or sensors and can, in principle, increase exponentially the overall capacities of such devices by putting them in network. The core mechanism of a QIN is entanglement swapping to propagate entanglement to the users' access point, who consume this entanglement resource to teleport the quantum states they wish to communicate. Quantum entanglement can be seen in this context as a new kind of network resource. QIN aims to distribute to end-users quantum information over long distances via quantum teleportation or entanglement swapping, trough ground and space paths, and enabling the realization of the quantum use case.

In this article, we presented architectures of QIN in which satellites can play a role for extending the range of communications. Satellites can embark entangled photons sources and/or Bell state measurement devices, acting as relay nodes for swapping the entanglement. We have distinguished two kinds of family architectures, depending on the use of quantum optical inter-satellite links. Four architecture types, with variations, have been discussed and compared. Only one architecture seems realistic for a short-term development and in-orbit demonstration: the entangled photon source on-board the satellite, in downlink configuration. The other architectures, involving Bell state measurement with space quantum memories, will demand much more R&D efforts.

## ACKNOWLEDGEMENTS

This work was supported by the French Space Agency, the Centre National d'Etudes Spatiales (CNES), the European Space Agency (ESA), and Thales Alenia Space.